\documentclass[epj,dvips]{webofc}
\usepackage[varg]{txfonts}   % Web of Conferences font
\providecommand{\tabularnewline}{\\}

\woctitle{Physics at the Magnetospheric Boundary}
\begin{document}
\title{MHD Simulations of Magnetospheric Accretion, Ejection and Plasma-field Interaction}

\author{M. M. Romanova\inst{1}\fnsep\thanks{\email{romanova@astro.cornell.edu}},
R. V. E. Lovelace\inst{1}\fnsep, M. Bachetti\inst{2}\fnsep, A. A.
Blinova\inst{1}\fnsep, A. V. Koldoba\inst{3,4}\fnsep,
 R. Kurosawa\inst{5}\fnsep,
   P. S. Lii\inst{1}\fnsep,
 \and
 G.~V. Ustyugova\inst{3},
}

\institute{
  Department of Astronomy, Cornell University, Ithaca, NY 14853-6801,
  USA
\and Institut de Recherche en Astrophysique et Planetologie, Toulouse, 31400,
  France
\and Keldysh Institute of Applied Mathematics, Moscow, 125047, Russia
 \and Moscow Institute of Physics and Technology, Dolgoprudny, Moscow Region, 141700,
 Russia
\and Max-Planck-Institut f\"{u}r Radioastronomie, Auf dem
  H\"{u}gel, 69, 5312 Bonn, Germany }

\abstract{\noindent We review recent axisymmetric and three-dimensional (3D)
magnetohydrodynamic (MHD) numerical simulations of magnetospheric accretion,
plasma-field interaction and outflows from the disk-magnetosphere boundary.}
\maketitle

\vspace{-0.3cm}
\section{Introduction}
\label{intro}

Dynamically important magnetic fields are present in many classes of accreting stars.
Some of them are very young, such as Classical T Tauri stars (CTTSs), while others are
very old --- white dwarfs and neutron stars. The light-curves from these stars show
complex patterns of periodic,
    quasi-periodic, or irregular variability which can be connected with different processes at
    the disk-magnetosphere boundary, such as complex paths of matter flow around the
    magnetosphere, rotations of the hot spots, waves excited in the inner
    disk, inflation and reconnection of external magnetic field lines of the magnetosphere,
    and other processes (e.g., \cite{BouvierEtAl2007}).
 Outflows are observed from a number of CTTSs (e.g., \cite{RayEtAl2007}) and a few accreting neutron
    stars and white dwarfs (e.g., \cite{Fender2004}). Some of the outflows may originate at the disk-magnetosphere boundary
    (e.g., \cite{ShuEtAl1994, FerreiraEtAl2006}).     The problem of the disk-magnetosphere
    interaction is multidimensional and requires global axisymmetric and three-dimensional (3D) numerical
    simulations. Below, we describe results of recent  numerical simulations of  magnetospheric
accretion and outflows from the disk-magnetosphere boundary.

\vspace{-0.3cm}

\section{Magnetospheric Accretion}
\label{sec:plasma-field} \vspace{-0.2cm}

%%%%%%% Figure 1 %%%%%%%%%%%%%%%%%%%%%%%%%%%%
\begin{figure*}
  \begin{center}
    \begin{tabular}{cc}
             \includegraphics[clip,height=0.24\textwidth]{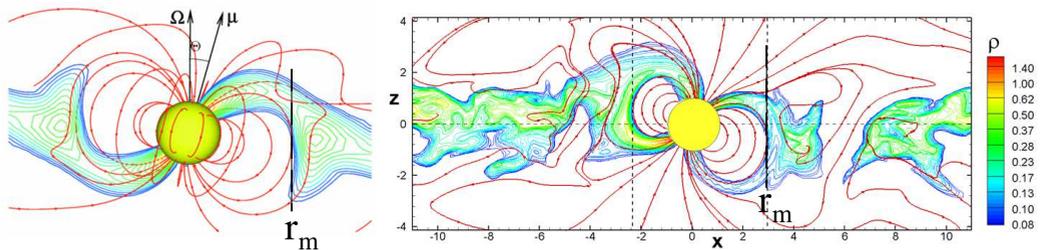}
    \end{tabular}
  \end{center}
\vspace{-0.7cm} \caption{Results of 3D MHD simulations of
accretion onto a star with a tilted dipole magnetic field in cases
of an $\alpha-$disk (left panel, from \cite{RomanovaEtAl2004}) and
a turbulent, MRI-driven disk (right panel, from
\cite{RomanovaEtAl2012}).} \label{funnel-2}
\end{figure*}
%%%%%%%%%%%%%%%%%%%%%%%%%%%%%%%%%%%%%%%%%%%%%

Magnetospheric accretion is a complex process, where the inner disk matter interacts
with the magnetosphere of the star. The result of such an interaction depends on a
number of factors, such as the size of the magnetosphere, the period of stellar
rotation, and the structure of the magnetic field of the star. Observational
properties of magnetized stars also depend on whether or not a star excites waves in
the inner disk. Significant progress has been achieved in understanding the
disk-magnetosphere interaction due to global 3D simulations performed with the
\textit{Cubed Sphere} code \cite{KoldobaEtAl2002, RomanovaEtAl2003}. In particular, it
was found that a magnetized star may accrete in either stable or unstable regimes (see
Sec. \ref{sec:stab-unstab}). It has also become possible to model accretion to stars
with complex magnetic fields and to model young stars with realistic parameters (Sec.
\ref{sec:complex}). In more recent 3D simulations, accretion to a star with a large
magnetosphere and from a very thin disk has been investigated (Sec.
\ref{sec:larger-magnetosphere}). It has also become possible to investigate waves in
the disk excited by a tilted rotating magnetosphere (Sec. \ref{sec:waves}).

\vspace{-0.3cm}
\subsection{Truncation of the disk by the magnetosphere}
\label{sec:magnetosphere} \vspace{-0.2cm}

If the magnetic field of the star is sufficiently strong, then it truncates the
accretion disk at a radius  $r_m$ where the magnetic stress in the magnetosphere
matches the matter stress in the disk. In the disk the main stress is connected with
the azimuthal components of the stress tensor: $T_{\phi\phi}=[p + \rho v_\phi^2] +
[B^2/8\pi-B_\phi^2/4\pi]$, where $\rho$, $p$, $B$, $B_\phi$ are local density, gas
pressure, total and azimuthal magnetic field in the disk (we neglect the viscous
stress, which is much smaller); $v_\phi$ is the azimuthal velocity in the reference
frame of the rotating magnetosphere:
 $v_\phi=v_{\rm disk}-v_{m}$.
 At the innermost edge of the disk, $B_\phi << B$ and the truncation radius is described by the condition
 $p + \rho v_\phi^2 = B^2/8\pi$.  We also introduce the
 modified plasma parameter

\vspace{-0.3cm}
\begin{equation}
\beta_1={8\pi}(p+\rho v^2)/{B^2}, \label{eq-beta1}
\end{equation}
\vspace{-0.3cm}

\noindent which is analogous to the standard plasma parameter $\beta=8\pi p/B^2$, but
the ram pressure of the disk matter $\rho v^2$ is included, where the total velocity
in a thin disk is $v\approx v_\phi$. Both axisymmetric and 3D simulations confirmed
that the inner disk truncates at the radius where $\beta_1\approx 1$ (e.g.,
\cite{RomanovaEtAl2002, LongEtAl2005, RomanovaEtAl2003, RomanovaEtAl2004}) . Fig.
\ref{funnel-2} shows examples of 3D simulations in the case of a laminar $\alpha-$disk
\cite{ShakuraSunyaev1973} (left panel, from \cite{RomanovaEtAl2004}) and a turbulent,
MRI-driven \cite{BalbusHawley1991} disk (right panel, from \cite{RomanovaEtAl2012}).
We noticed that the condition $\beta=8\pi p/B^2=1$ can also be used, in particular, in
cases where $v_{\rm disk}\approx v_m$, and also for finding the position of the funnel
streams, which are at slightly larger distances than the inner edge of the disk.
\cite{BessolazEtAl2008} derived the magnetospheric radius from different conditions,
and found only small difference between different radii $r_m$.

 The magnetospheric
radius has also been derived theoretically in analogy with the Alfv\'en radius in the
case of spherical accretion (e.g., \cite{LambEtAl1973}):

\vspace{-0.3cm}
\begin{equation}
r_m = k \big[\mu^4/(\dot{M}^2 GM_\star)\big]^{1/7}, ~~~~~k\sim 1~, \label{eq-alfven}
\end{equation}
\vspace{-0.3cm}

\noindent where $\mu=B_\star R_\star^3$ is the magnetic moment of the star with a
surface field $B_\star$, $\dot{M}$ is the disk accretion rate, $M_\star$  and
$R_\star$ are the mass and radius of the star. This formula has been tested in a few
axisymmetric simulations by \cite{LongEtAl2005, ZanniFerreira2013}. It was found that
eqs. \ref{eq-beta1} and \ref{eq-alfven} give the same $r_m$ if $k\approx 0.5$
\cite{LongEtAl2005}. \cite{KulkarniRomanova2013} re-investigated this issue in
multiple 3D simulations while taking $\beta=1$ for finding $r_m$. It was found that
the dependence is slightly different: ${r_m}/R_\star \approx 1.06 \large[
{\mu^4}/(\dot{M}^2 GM_\star R_\star^7)\large]^{1/10} ,$ which can be explained by the
compression of the magnetosphere and its departure from the dipole shape.

%%%%%%%% Figure 2 %%%%%%%%%%%%%%%%%%%%%%%%%%%%
\begin{figure*}
  \begin{center}
    \begin{tabular}{cc}
             \includegraphics[clip,height=0.32\textwidth]{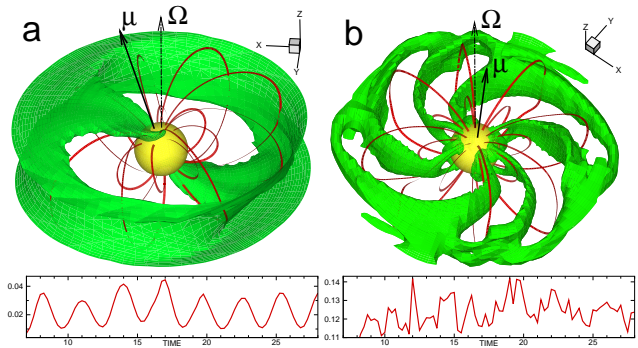}\hspace{0.0cm}
    \end{tabular}
  \end{center}
\vspace{-0.7cm} \caption{\textit{Left panel:} a 3D view of matter
flow in the stable regime of accretion. \textit{Right panel:} a 3D
view of accretion in the unstable regime. Bottom panels show the
light-curves from the hot spots. From \cite{RomanovaEtAl2008}.
 }
\label{3d-stab-unstab}
\end{figure*}
%%%%%%%%%%%%%%%%%%%%%%%%%%%%%%%%%%%%%%%%%%%%%

\vspace{-1.2cm}
\subsection{Stable and unstable regimes of accretion}
\label{sec:stab-unstab}

\vspace{-0.2cm}

 Global 3D simulations of accretion to stars with relatively small magnetospheres
 (a few stellar radii) show that accreting magnetized stars may be
 in either stable or unstable regimes of accretion. In the stable regime,
matter accretes to the star in two ordered funnel streams and
    forms two antipodal spots at its surface \cite{KoldobaEtAl2002, RomanovaEtAl2003,
    RomanovaEtAl2004}.  Rotation of the star gives an almost sinusoidal light-curve (see Fig. \ref{3d-stab-unstab}, left panel).
    Such light-curves are observed, e.g., in accreting millisecond pulsars (AMPs) \cite{IbragimovPoutanenEtAl2009}
    and many CTTSs \cite{HerbstEtAl1994}.
    In the unstable regime, matter penetrates
    through the magnetosphere in chaotic equatorial  ``tongues" due to the
    magnetic Rayleigh-Taylor (R-T) instability \cite{RomanovaEtAl2008,
    KulkarniRomanova2008}, and the light-curve from the hot spots may be chaotic (see
    Fig. \ref{3d-stab-unstab}, right panels). Chaotic light-curves with no clear periods are observed in many CTTSs \cite{HerbstEtAl1994}.
    In the intermediate regime, both the magnetospheric funnel
    and the unstable tongues are observed \cite{KulkarniRomanova2009, BachettiEtAl2010}.
Earlier, \cite{AronsLea1976} suggested that the R-T and Kelvin-Helmholtz instabilities
may be efficient mechanisms of mixing accreting matter to magnetosphere (see also
\cite{LovelaceEtAl2010a}). Simulations show that the unstable regime may lead to a
qualitatively different state of accretion.

Observational properties of CTTSs in stable and unstable regimes have been studied by
\cite{KurosawaEtAl2008, KurosawaRomanova2013}, where the cubed sphere grid of the 3D
MHD code has been projected to the adaptive mesh refinement grid of 3D radiative
transfer code $\sc Torus$ \cite{Harries2000}, and time-dependent hydrogen spectral
lines were calculated from 3D MHD modeled flow. These 3D+3D simulations have shown
that in the stable regime, periodic obscuration of stellar light by a funnel stream
(where matter moves away from the observer) produces periodic variability of the
redshifted wing of spectral lines, while in the unstable regime (where several tongues
are usually present in the field of view of the observer), the redshifted absorption
is quasi-steady.

The boundary between stable and unstable regimes  has been investigated by
\cite{RomanovaEtAl2008, KulkarniRomanova2008}. They found that at a fixed period of
the star, the boundary depends on the accretion rate: accretion becomes unstable at
higher accretion rates. The accretion rate has been varied using the
$\alpha-$parameter of viscosity \cite{ShakuraSunyaev1973}. In more recent simulations,
performed at a higher grid resolution and a small $\alpha=0.02$
\cite{BlinovaEtAl2013a}, the period of the star has been varied, and a simple
criterion has been derived: the accretion is unstable if $r_c \gtrsim 1.4 r_m$ , where
$r_c =(GM_\star/\Omega_\star^2)^{1/3}$ is the corotation radius, and $\Omega_\star$ is
the angular velocity of the star. In both investigations, the boundary between the
regimes agrees with the theoretical study by \cite{SpruitEtAl1995}.

In the cases of very small magnetospheres, $r_m\lesssim (1-2) R_\star$, unstable
accretion becomes more ordered, with one or two unstable tongues forming and rotating
with the angular velocity of the inner disk. This may cause the frequency of the inner
disk to dominate in the spectrum \cite{RomanovaKulkarni2009}.  The typical feature of
this frequency is that it varies in time: it increases with accretion rate, when the
inner disk moves inwards. In CTTSs, this frequency can be mistaken for the frequency
of the star, or may lead to the phenomenon of a drifting ``period", observed in some
CTTSs (e.g., \cite{RucinskiEtAl2008}). In AMPs, two high-frequency QPOs are observed
\cite{vanderKlis2006}. The upper frequency, $\nu_u$, may be connected with the
rotation of the unstable tongues. The lower frequency, $\nu_l$, may be connected with
the rotation of a funnel stream about the magnetic pole, which is possible in stars
with very small tilts of the magnetosphere \cite{BachettiEtAl2010}.
 Both the funnels and the tongues produce hot spots on the surface of the star which rotate
 with similar frequencies. Simulations show that both frequencies vary strongly with accretion
 rate.
 However, their difference, $\nu_u-\nu_l$, does not vary much and
 does not correlate with the frequency of the star,
 which is typical for a number of AMPs  \cite{BelloniEtAl2007}. The phenomenon of rotating funnels and tongues
is probably determined or enhanced by waves in the inner disk
 (see Sec. \ref{sec:waves}).

%%%%%%% Figure 3 %%%%%%%%%%%%%%%%%%%%%%%%%%%%
\begin{figure*}
  \begin{center}
    \begin{tabular}{cc}
             \includegraphics[clip,height=0.3\textwidth]{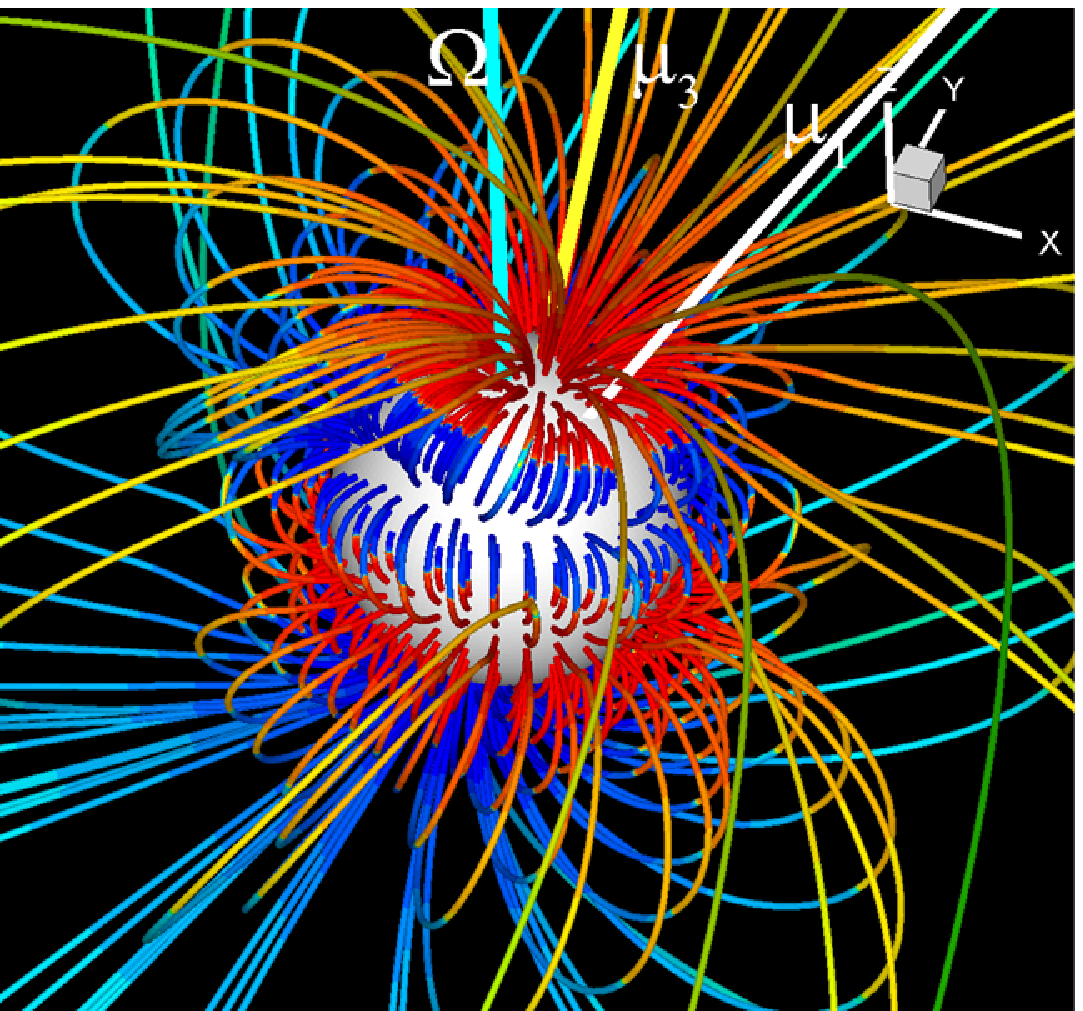}\hspace{0.3cm}
             &
      \includegraphics[clip,height=0.3\textwidth]{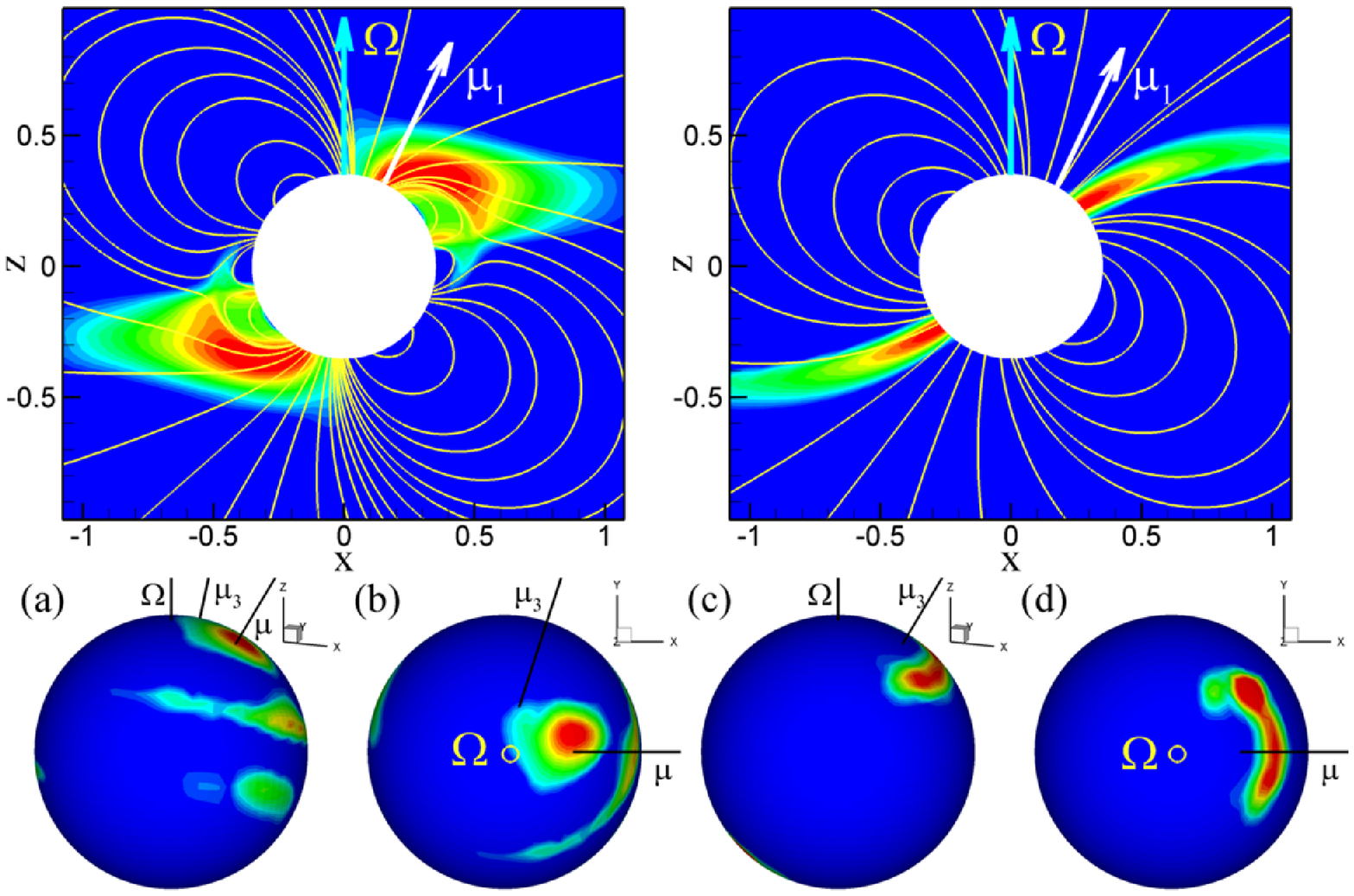}\tabularnewline
\end{tabular}
  \end{center}
\vspace{-0.7cm} \caption{{\it Left panel:} a 3D view of the
initial magnetic field in a 3D MHD model of a young star V2129
Oph, where the magnetic field is dominated by a 0.35 kG dipole and
1.2 kG octupole components.  {\it Right panel:} slices of the
density distribution and hot spots (energy distribution at the
surface of the star) in cases of the dipole+octupole field and a
pure dipole field.  From \cite{RomanovaEtAl2011a}.}
\label{magn-large}
\end{figure*}
%%%%%%%%%%%%%%%%%%%%%%%%%%%%%%%%%%%%%%%%%%%%%

\vspace{-0.3cm}
\subsection{Accretion to stars with complex magnetic field}
\label{sec:complex} \vspace{-0.2cm}

 Stars may have a complex
magnetic field (e.g., \cite{DonatiEtAl2007,JohnsKrull2007,Gregory2011}). The magnetic
field can be decomposed into a superposition of tilted dipole, quadrupole, octupole
and higher order multipoles: $\bf B_\star =  {\bf B}_{\rm dip} + {\bf B}_{\rm quad} +
{\bf B}_{\rm oct} + ...$.
% and complex paths of matter flow are expected (e.g., \cite{Gregory2011}).
Global 3D simulations show that if the magnetic field of a star has dipole and
quadrupole field components, then matter partially accretes to the regions of the
magnetic poles, and partially to the ring associated with the quadrupole component of
the field \cite{LongEtAl2007, LongEtAl2008}. In the case of a superposition of dipole
and octupole fields, matter partially flows to two octupole rings \cite{LongEtAl2012}.
It is often the case that the dipole component dominates at large distances from the
star and determines the disk-magnetosphere interaction and the initial path of the
funnel streams, while the quadrupole or octupole components determine matter flow
close to the star and influence the shapes of the hot spots. Recent measurements of
the surface magnetic field in CTTSs have shown that in several stars the dominant
components of the field are dipole and octupole, while quadrupole and higher-order
components are smaller \cite{DonatiEtAl2007, DonatiEtAl2008}. This helped us model the
magnetospheric accretion to CTTSs with realistic parameters: V2129 Oph, where the
octupole component is relatively large \cite{RomanovaEtAl2011a}, and BP Tau, where the
octupole component is small \cite{LongEtAl2011}. Hot spots derived from simulations
were in good agreement with the spots obtained from polarimetric observations and
analyses \cite{DonatiEtAl2007, DonatiEtAl2008}.

More recently, a time-variable spectrum in hydrogen lines has been
calculated from the 3D MHD model of V2129 and has been compared
with the time-variable spectrum obtained from the observations of
this star \cite{AlencarEtAl2012}. Comparisons show a good match
between the observed and modeled spectra.

\vspace{-0.3cm}
\subsection{Larger magnetosphere, thinner disk}
\label{sec:larger-magnetosphere}
\vspace{-0.2cm}

 Earlier simulations are relevant to CTTSs and AMPs, where the size of the magnetosphere
 is a few times larger than the radius of the star. However, in intermediate polars
    $r_m/R_\star \gtrsim 10$, while in X-ray pulsars $r_m$ may reach hundreds of stellar
    radii.
 More recent simulations were aimed at investigating the
disk-magnetosphere interaction in cases of larger magnetospheres, where the
magnetospheric radius $r_m>10 R_\star$ (Romanova et al., in prep). Preliminary
simulations show that matter of the inner disk penetrates
    through the external layers of the magnetosphere up to the depth of 1-2 stellar radii
(see Fig. \ref{magn-large}). Later on, a few funnel streams form in regions of
enhanced density, which coincide with unstable tongues.  It is often the case that a
few funnel streams form because a few unstable tongues are located in the region of
the disk where accretion through funnel streams is favorable. In current simulations,
we see that  accretion through R-T instability does not lead to the globally-chaotic
accretion, as in the case of small magnetospheres.  However, this result should be
checked+ at a wider range of parameters.

 In another set of numerical experiments, the grid was compressed towards the
 disk, and accretion from a very thin disk ($h/r\approx 0.02-0.03$) was investigated.
 The main goal was to understand how matter of the inner disk ``climbs" to the magnetospheric wall from the
equatorial plane without a
 sufficient pressure gradient or other forces lifting matter above the equatorial
 plane. Simulations were done for a relatively small magnetosphere of $\sim 5 R_\star$, and
the tilt of the dipole was $\Theta=20^\circ$. Simulations have
shown an interesting phenomenon. The tilted dipole excited bending
waves in the disk, which helped lift matter from the equatorial
plane to larger heights, from where the gravitational force pulled
matter towards the star. This mechanism of lifting may be
important in cases of much larger magnetospheres, as in
intermediate polars and X-ray pulsars.

%%%%%%% Figure 3 %%%%%%%%%%%%%%%%%%%%%%%%%%%%
\begin{figure*}
  \begin{center}
    \begin{tabular}{cc}
             \includegraphics[clip,height=0.22\textwidth]{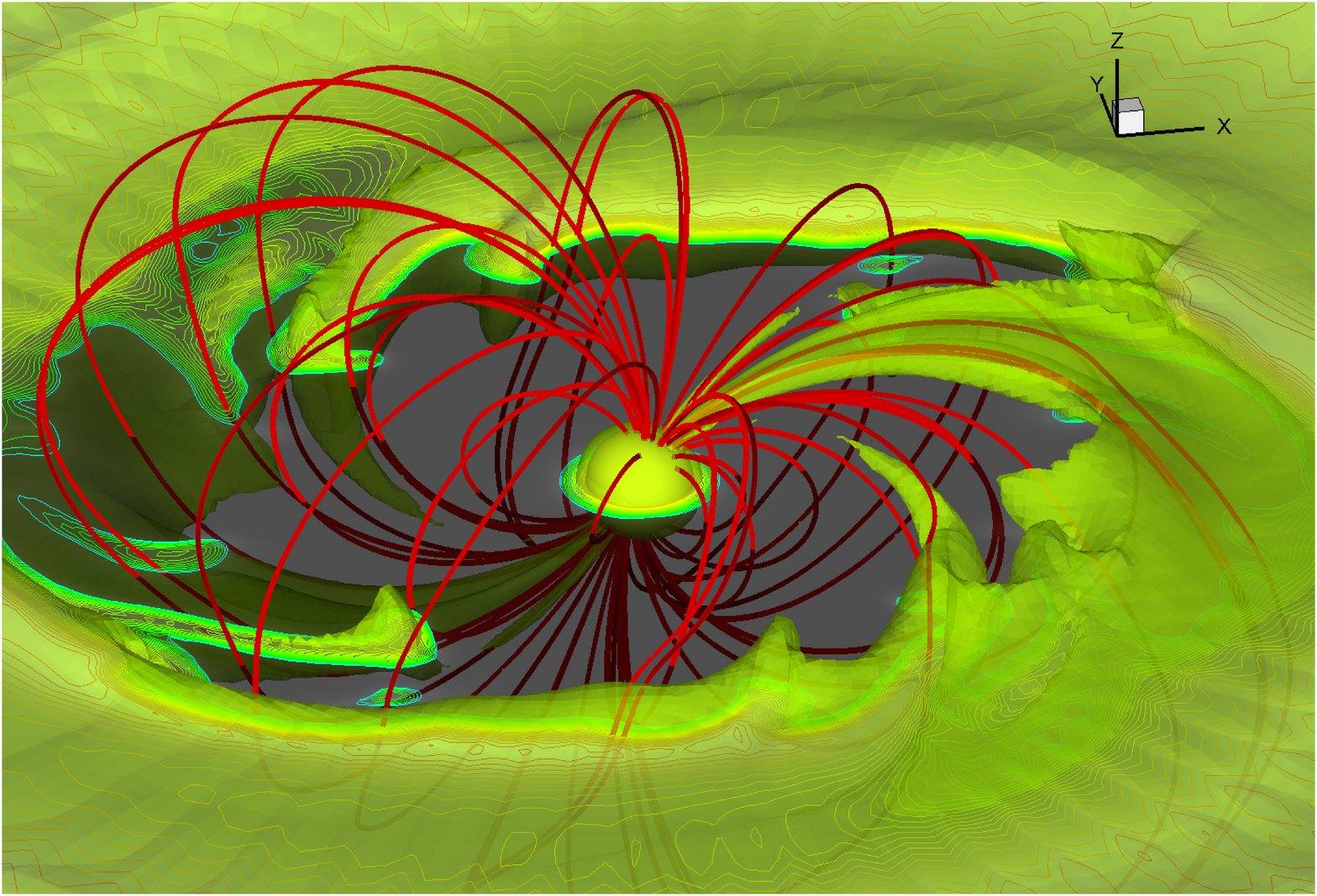}\hspace{0.3cm}
             &
      \includegraphics[clip,height=0.22\textwidth]{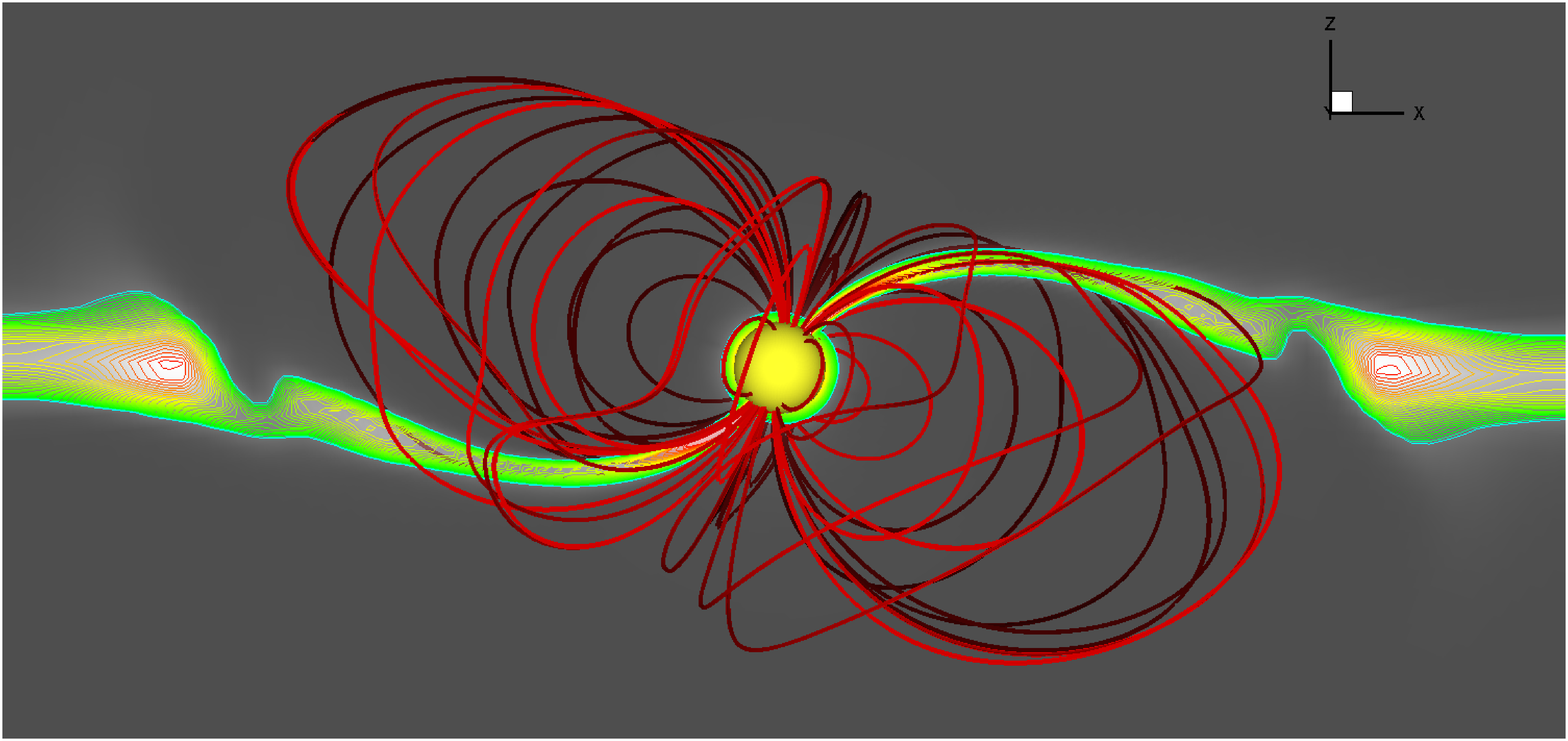}\tabularnewline
\end{tabular}
  \end{center}
\vspace{-0.7cm} \caption{{\it Left panel:} a 3D view of matter
flow in the case of a star with a relatively large magnetosphere
($r_m\approx 12 R_\star$). {\it Right panel:} an XZ-slice of
density distribution and sample field lines.} \label{magn-large}
\end{figure*}
%%%%%%%%%%%%%%%%%%%%%%%%%%%%%%%%%%%%%%%%%%%%%

\vspace{-0.3cm}
\subsection{Warps and waves in the disk}
\label{sec:waves}

 \vspace{-0.2cm}
A tilted magnetosphere applies force to the inner parts of the disk and excites waves
in the disk \cite{LipunovShakura1980, Lai1999}. If the rotational axis of the star is
aligned with the rotational axis of the disk, then waves in the disk are excited on
the time-scale of stellar rotation. Global 3D simulations show that when the
magnetosphere and the inner disk are in approximate corotation (that is, $r_m\approx
r_c$), a large bending wave is formed and rotates with the angular velocity of the
star \cite{RomanovaEtAl2013}. The height of the wave depends on the density level of
the warp and is approximately $(0.1-0.2) R_\star$. This result is in agreement with
the theoretical findings of \cite{TerquemPapaloizou2000}. This wave may obscure the
light of the star and lead to a special type of variability, with regular dips in the
light-curves of CTTSs \cite{BouvierEtAl1999, AlencarEtAl2010}. In the cases where the
magnetosphere rotates more slowly than the inner disk ($r_m<r_c$), two types of waves
are observed in the inner disk. One is  a radially-trapped density wave, which appears
in the region of the maximum of angular
 velocity in the inner disk, where its rotation rate varies from Keplerian to that of the star
 \cite{LovelaceRomanova2007}. The other one is an inner bending wave which rotates with
near-Keplerian velocity. Both frequencies depend on the position of the inner disk and
vary with accretion rate \cite{RomanovaEtAl2013}.  These two waves may also explain
the twin-peak QPOs observed in AMPs: the trapped density wave may be responsible for
the strongest (lower) QPO frequency, while the inner bending wave may be responsible
for the upper QPO. Again, the difference $\nu_u-\nu_l$ does not correlate with the
frequency of the star.

\begin{figure*}
  \begin{center}
    \begin{tabular}{cc}
\includegraphics[clip,height=0.24\textwidth]{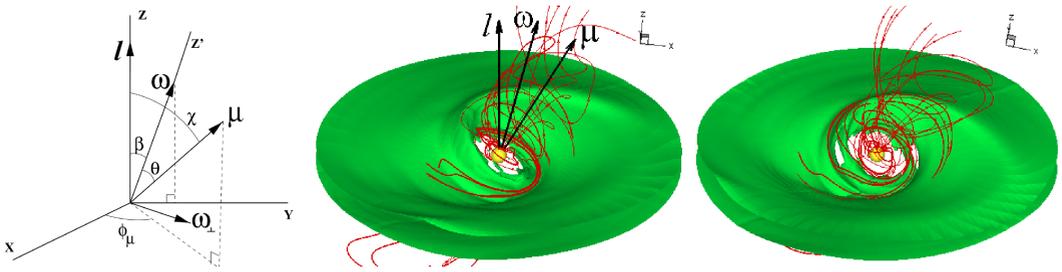}
             \tabularnewline
    \end{tabular}
  \end{center}
\vspace{-0.7cm} \caption{Results of 3D simulations of disk accretion to a star where
both the rotational ($\omega$) and the magnetic ($\mu$) axes are tilted with respect
to the rotational axis of the disk. } \label{3d-tilted-3}
\end{figure*}

If the rotational axis of the star is tilted with respect to the rotational axis of
the disk, then the magnetic force acts systematically on the inner disk, pushing it to
tilt and to precess \cite{LipunovShakura1980, Lai1999}. Experimental numerical
simulations (Romanova et al., in prep) show that the inner disk is tilted and
precesses slowly (see Fig. \ref{3d-tilted-3}). In this case, the warp has a high
amplitude and can obscure or reflect light of the star on the time-scale of the
precession. Simulations also show that such a tilt of the inner disk influences the
magnetospheric accretion, in particular, the location of the funnel streams and hot
spots on the surface of the star. A precessing inner disk may also reflect light from
the star and provide low-frequency oscillations, or a modulation of light with the
frequency of the precession. This mechanism may explain some of the low-frequency
oscillations observed in different magnetized stars.

\vspace{-0.3cm}
\section{Outflows from the disk-magnetosphere boundary} \label{sec:outflows}

\vspace{-0.2cm}

At the disk-magnetosphere boundary, the conditions may be favorable for driving
outflows. Matter of the inner disk may be redirected to the outflows if the
centrifugal and/or magnetic pressure forces overcome the gravitational force and the
magnetic tension force of the closed magnetic field lines of the magnetosphere. The
differential rotation of the field lines connecting the star and the disk often leads
to inflation and opening of these field lines, which is favorable for outflows.
Different regimes of outflows are possible. If the magnetosphere rotates more rapidly
than the inner disk (that is, $r_m>r_c$), then the star is in the propeller regime
\cite{IllarionovSunyaev1975, LovelaceEtAl1999}, where the centrifugal force may drive
outflows. On the other hand, if the star rotates more slowly than the inner disk, then
the outflows may be driven by the magnetic force. Here, we consider these two types of
outflows.

%%%%%%% Figure 5 %%%%%%%%%%%%%%%%%%%%%%%%%%%%
%%%%%%%%%%%%%%%%%%%%%%%%%%%%%%%%%%%%%%%%%%%%%

\vspace{-0.3cm}
\subsection{Propeller-driven Winds and Jets}
\vspace{-0.2cm}

\begin{figure*}
  \begin{center}
    \begin{tabular}{cc}
\includegraphics[clip,height=0.27\textwidth]{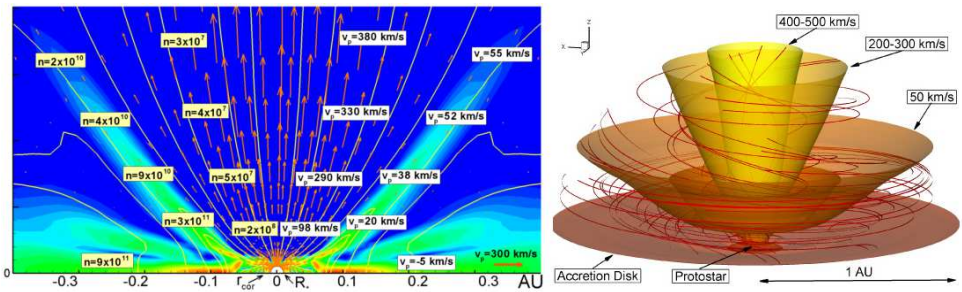}
             \tabularnewline
    \end{tabular}
  \end{center}
\vspace{-0.7cm} \caption{\textit{Left panel:}  density and velocity distribution in
the propeller regime with numbers corresponding to CTTSs \cite{RomanovaEtAl2009}.
\textit{Right Panel:} 3D rendering shows magnetic field lines and density levels
corresponding to different velocities. } \label{propeller-onion-2}
\end{figure*}

Axisymmetric simulations show that in the propeller regime a significant fraction of
the disk mass
%flow
can be redirected to the outflows by the rapidly-rotating magnetosphere
\cite{RomanovaEtAl2005, RomanovaEtAl2009, UstyugovaEtAl2006}. In this regime,
accretion and outflows occur in cycles, where matter accumulates in the inner disk,
diffuses across the field lines of the rapidly-rotating magnetosphere, and is ejected
to the outflows; then, the magnetosphere expands and the cycle repeats (see also
\cite{GoodsonEtAl1997, DangeloSpruit2010, LiiEtAl2013}). Most of the matter flows to a
conically-shaped wind, which is only gradually collimated with distance
\cite{RomanovaEtAl2009, LiiEtAl2013} (see Fig. \ref{propeller-onion-2}). In addition,
a smaller amount of matter flows to a magnetically-dominated, well-collimated Poynting
jet \cite{LovelaceEtAl2002}, where matter is accelerated rapidly by the magnetic force
\cite{LovelaceEtAl1991}. The jet carries significant energy and angular momentum out
of the star, causing it to spin down (e.g., \cite{RomanovaEtAl2005}). Part of the
angular momentum is transferred from the star to the partially-inflated field lines
\cite{ZanniFerreira2013, LiiEtAl2013}. In CTTSs, the propeller mechanism may be
responsible for their spinning down from near break-up speed to less than 10\% of this
speed (e.g., \cite{RomanovaEtAl2005}). In AMP SAX J1808.4-3658,  1Hz flaring
oscillations have been observed at the end of an outburst \cite{vanStraatenEtAl2005},
 and may be connected with the propeller regime
\cite{PatrunoEtAl2009,PatrunoDangelo2013}. A similar phenomenon has been observed in a
few cataclysmic variables, e.g., in AE Aqr (e.g., \cite{Mauche2006}).

\vspace{-0.3cm}
\subsection{Conical Winds}
\vspace{-0.2cm}

Recent numerical simulations show a new type of winds which can be important during
episodes of enhanced accretion rate, as during the outbursts of accretion in AMPs, or
in young EXOR and FUOR type stars \citep{RomanovaEtAl2009}. Simulations show that the
newly-incoming matter compresses the magnetosphere of the star, the field lines
inflate due to the differential rotation between the disk and the star, and
conically-shaped winds flow out of the inner disk \cite{RomanovaEtAl2009,
KurosawaRomanova2012}. These winds are driven  by the magnetic force, $F_M\propto -
\nabla (rB_\phi)^2$, which arises due to the wrapping of the field lines above the
disk \cite{LovelaceEtAl1991}. The wind is also gradually collimated by the magnetic
hoop-stress, and can be strongly collimated in the cases of high accretion rates
\cite{LiiEtAl2012}. A rapid rotation of the star is not required. Moreover, the star
can rotate much more slowly than the inner disk (at $r_m<<r_c$), which may be typical
during accretion outbursts in different types of stars. This is different from the
X-winds, which require the condition $r_m\approx r_c$. Conical winds may appear during
an outburst of accretion and continue for the entire duration of the outburst. A
magnetic field of a few kG is required for FUORs, while in EXORs a and CTTSs the field
can be weaker. The conical wind model has been applied to FU Ori and compared with the
empirical model based on the spectral analysis of the winds in FU Ori
\cite{CalvetEtAl1993}. A reasonably good agreement was found between these models
\cite{KoniglEtAl2011}.

\begin{figure*}[!ht]
  \begin{center}
    \begin{tabular}{cc}
\includegraphics[clip,height=0.24\textwidth]{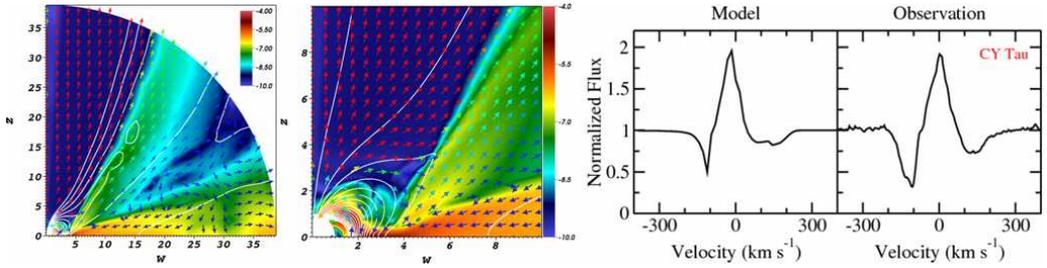}
             \tabularnewline
    \end{tabular}
  \end{center}
\vspace{-0.7cm} \caption{\textit{Left two panels:} axisymmetric
simulations of conical winds, shown on larger and smaller scales.
\textit{Right two panels:} a comparison of modeled and observed
spectra in CTTS CY Tau. From \cite{KurosawaRomanova2012}.}
\label{conical-spec-4}
\end{figure*}

Observational properties of conical winds in application to CTTSs were investigated in
\cite{KurosawaEtAl2011, KurosawaRomanova2012}, where the spectrum in He and H lines
was calculated using the  {\it Torus} code. Simulations show that conical winds
produce a narrow blue absorption component in the spectrum (see Fig.
\ref{conical-spec-4}). Such a blue component is frequently observed in the spectra of
CTTSs (e.g., \cite{EdwardsEtAl2006}).

%%%%%%% Figure 7 %%%%%%%%%%%%%%%%%%%%%%%%%%%%
%%%%%%%%%%%%%%%%%%%%%%%%%%%%%%%%%%%%%%%%%%%%%

%%%%%%% Figure 7 %%%%%%%%%%%%%%%%%%%%%%%%%%%%
%%%%%%%%%%%%%%%%%%%%%%%%%%%%%%%%%%%%%%%%%%%%%

\vspace{-0.3cm}
\subsection{Asymmetric and one-sided outflows}

\vspace{-0.2cm}

\begin{figure*}[!ht]
  \begin{center}
    \begin{tabular}{cc}
\includegraphics[clip,height=0.3\textwidth]{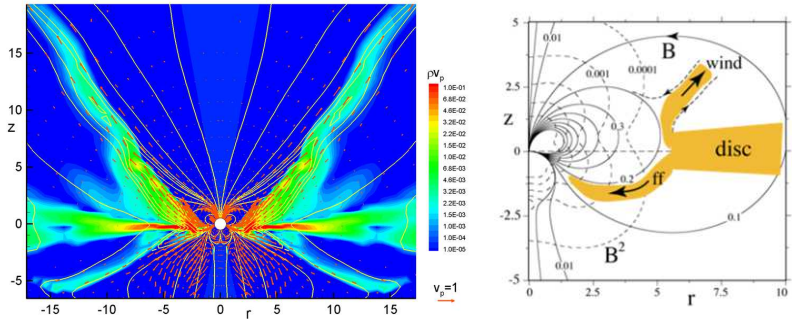}
             \tabularnewline
    \end{tabular}
  \end{center}
\vspace{-0.7cm} \caption{\textit{Left panel:} a density slice and sample field lines
show the result of an axisymmetric simulation of accretion to a star with dipole and
quadrupole fields. \textit{Right panel:} a sketch of accretion and outflows (from
\cite{LovelaceEtAl2010b}).}\label{asym-2}
\end{figure*}

In stars with a complex magnetic field, outflows may be asymmetric
due to the top-bottom asymmetry of the magnetic field. For
example, the superposition of an axisymmetric dipole and a
quadrupole leads to a magnetic configuration where the magnetic
flux is larger on one side of the equatorial plane and smaller on
the other side (see Fig. \ref{asym-2}, right panel). Axisymmetric
simulations of the propeller regime show that stronger outflows
are observed on the side where the magnetic flux is larger (see
left panel of Fig. \ref{asym-2}). In this case, the matter and
energy fluxes will be systematically higher in one direction, and
lower in the other direction. One-sided outflows are observed in a
number of young stars \cite{BacciottiEtAl1999}.

Axisymmetric simulations of the entire region also show that even in the case of a
pure dipole field, outflows are usually one-sided. However, the direction of the
outflows switches frequently, and therefore the averaged matter and energy fluxes of
the outflows above and below the equatorial plane are expected to be approximately
equal in both directions \cite{LovelaceEtAl2010b}. Recent simulations of the propeller
regime in the case of MRI-driven accretion have also shown that the outflows are
one-sided, but the wind switches sides much less frequently \cite{LiiEtAl2013}.

%%%%%%% Figure 8 %%%%%%%%%%%%%%%%%%%%%%%%%%%%
%%%%%%%%%%%%%%%%%%%%%%%%%%%%%%%%%%%%%%%%%%%%%

\vspace{-0.5cm}
\section{Disk-magnetosphere interaction and variability}

\vspace{-0.2cm}

Here, we briefly discuss the different variabilities expected during the
disk-magnetosphere interaction.  In the \textit{stable regime}, the hot spots rotate
approximately with the period of the star, although variation of the accretion rate
leads to phase-shifts of spots in the azimuthal direction: the phase increases with
accretion rate \cite{RomanovaEtAl2004, RomanovaEtAl2013}. This may lead to the
phenomenon of a drifting period in CTTSs, where the period is often derived from the
observations of hot spots. In AMPs, the variation of phase with luminosity has been
observed (e.g., \cite{PapittoEtAl2007}). Variation of phase may lead to the observed
gradual variation of the ``period" of the star, which is determined from the
observations of hot spots. In the \textit{unstable regime}, accretion is often
stochastic, with a time-scale of a few accretion events per period of the inner disk.
However, the entire set of unstable tongues rotates with the angular velocity of the
inner disk, and this frequency may be seen in the power spectra. The period of the
star can also be seen, unless the instability is very strong
\cite{KulkarniRomanova2009, BachettiEtAl2010}. In the cases of very small
magnetospheres, $r_m\lesssim 2 R_{\star}$, one or two unstable tongues dominate and
rotate with the \textit{frequency of the inner disk} \cite{RomanovaEtAl2009}.  In the
case of CTTSs, this period can be mistakenly interpreted as the strongly drifting
``period" of the star. The rotation of the foot-points
 of moving funnel streams and unstable tongues may be determined or enhanced by
 waves in the inner disk.

 Longer time variability (a few periods of the inner disk or more) can be connected
 with the
 \textit{opening and reconnection} of the external field lines of the magnetosphere
 \cite{AlyKuijpers1990}. This phenomenon has been observed in the conical wind model
 \cite{RomanovaEtAl2009, LiiEtAl2012} and in the Magnetospheric Ejections model
 \cite{ZanniFerreira2013}. These oscillations may have very long periods in the cases of
 very low diffusivity. If the rotational axes of the star and the disk are \textit{misaligned}, then
a slow precession of the inner disk is expected (\cite{Lai1999},
Sec. \ref{sec:waves}), and this may be a source of persistent
low-frequency oscillations.
 Another type of low-frequency oscillations may be connected
 with episodic accumulation and accretion of matter in a \textit{weak propeller regime} \cite{DangeloSpruit2010}.

\section*{Acknowledgments}
\vspace{-0.2cm}

 We thank the conference organizers, especially Dr. E. Bozzo,
for the excellent meeting. MMR is grateful for the generous
support of her trip. Resources supporting this work were provided
by the NASA High-End Computing (HEC).  The research was supported
in part by NASA grant NNX11AF33G and NSF grant AST-1211318. GVU
and AVK
 were supported in part by FAP-14.B37.21.0915, SS-1434.2012.2, and RFBR
12-01-00606-a.

\bibliography{local}

\vspace{-0.3cm}

\begin{thebibliography}{}

\vspace{-0.2cm}


\bibitem{BouvierEtAl2007} Bouvier, J.,  Alencar, S. H. P.,
Harries, T. J., Johns-Krull, C. M., Romanova, M. M.,  \textit{Protostars and Planets}
V, B. Reipurth, D. Jewitt, and K. Keil (eds.), (University of Arizona Press, Tucson,
2007), p. 479


\bibitem{RayEtAl2007} Ray, T., Dougados, C., Bacciotti, F., Eisl\"offel, J.,
 Chrysostomou, A., \textit{Protostars and Planets} V, B. Reipurth, D. Jewitt,
 and K. Keil (eds.), (University of Arizona Press, Tucson, 2007) p. 231


\bibitem{Fender2004} Fender, R.P., Nature \textbf{427}, 222 (2004)


\bibitem{ShuEtAl1994} Shu, F., Najita, J., Ostriker, E., Wilkin, F., Ruden, S.,
Lizano, S., ApJ \textbf{429}, 781 (1994)

\bibitem{FerreiraEtAl2006} Ferreira, J., Dougados, C., Cabrit, S., A\&A
\textbf{453}, 785 (2006)


\bibitem{KoldobaEtAl2002} Koldoba, A. V., Romanova, M. M., Ustyugova, G. V., Lovelace, R. V. E.,
ApJ \textbf{576}, L53 (2002)

\bibitem{RomanovaEtAl2003} Romanova, M.M., Ustyugova, G.V., Koldoba, A.V.,
Lovelace, R.V.E. 2003, ApJ \textbf{595}, 1009 (2003)


\bibitem{RomanovaEtAl2002} Romanova, M.M., Ustyugova, G.V., Koldoba, A.V., \&
Lovelace, R.V.E., ApJ \textbf{578}, 420 (2002)


\bibitem{LongEtAl2005} Long, M., Romanova, M.M.,  Lovelace, R.V.E., ApJ
\textbf{634}, 1214 (2005)



\bibitem{RomanovaEtAl2004} Romanova, M.M., Ustyugova, G.V., Koldoba, A.V.,
Lovelace, R.V.E., ApJ \textbf{610}, 920 (2004)


\bibitem{ShakuraSunyaev1973} Shakura, N.I., \& Sunyaev, R.A., A\&A \textbf{24}, 337 (1973)


\bibitem{BalbusHawley1991}   Balbus, S.A., \& Hawley, J.F., ApJ \textbf{376}, 214 (1991)


\bibitem{RomanovaEtAl2012} Romanova, M.M., Ustyugova, G.V., Koldoba, A.V.,
Lovelace, R.V.E., MNRAS \textbf{421}, 63 (2012)


\bibitem{BessolazEtAl2008} Bessolaz N., Zanni C., Ferreira J., Keppens R., Bouvier J., A\&A \textbf{478},
155 (2008)


\bibitem{LambEtAl1973} Lamb, F.~K., Pethick, C.~J., Pines, D., ApJ, \textbf{184}, 271
(1973)


\bibitem{ZanniFerreira2013} Zanni, C., \& Ferreira, J., A\&A \textbf{550}, A99 (2013)


\bibitem{KulkarniRomanova2013} Kulkarni, A., \& Romanova, M.M., MNRAS \textbf{433},
3048 (2013)


\bibitem{IbragimovPoutanenEtAl2009} Ibragimov, A., \& Poutanen, J., MNRAS
\textbf{400}, 492

\bibitem{HerbstEtAl1994} Herbst, W., Herbst, D. K., Grossman, E. J., Weinstein, D. 1994, AJ \textbf{108},
1906


\bibitem{RomanovaEtAl2008} Romanova, M.M., Kulkarni, A.K., Lovelace, R.V.E.,
ApJ \textbf{673}, L171 (2008)


\bibitem{KulkarniRomanova2008} Kulkarni, A., \& Romanova, M.M., MNRAS \textbf{386},
673 (2008)



\bibitem{KulkarniRomanova2009} Kulkarni, A., \& Romanova, M.M., MNRAS \textbf{386}, 673
(2009)

\bibitem{BachettiEtAl2010}  Bachetti, M., Romanova, M.M., Kulkarni, A.,
 Burderi, L., di Salvo, T., MNRAS \textbf{403}, 1193 (2010)


\bibitem{AronsLea1976} Arons, J., \& Lea, S.M., ApJ \textbf{207}, 914 (1976)



\bibitem{LovelaceEtAl2010a} Lovelace, R.V.E., Romanova, M.M., Newman, W.I., MNRAS,
\textbf{402}, 2575 (2010)


\bibitem{KurosawaEtAl2008} Kurosawa, R, Romanova, M.M., Harries, T.J., MNRAS \textbf{385}, 1931 (2008)


\bibitem{KurosawaRomanova2013} Kurosawa, R., \& Romanova, M.M., MNRAS \textbf{431},
2673 (2013)


\bibitem{Harries2000} Harries, T.J.,  MNRAS \textbf{315}, 722 (2000)


\bibitem{BlinovaEtAl2013a} Blinova, A.A., Romanova, M.M., \& Lovelace, R.V.E.,
  \textit{Physics at the Magnetspheric Boundary} (Geneva,
 Switzerland, 2013) (arXiv:1309.4363)


\bibitem{SpruitEtAl1995}
Spruit H. C., Stehle R., Papaloizou J. C. B., MNRAS \textbf{275}, 1223 (1995)


\bibitem{RomanovaKulkarni2009} Romanova, M.M., \& Kulkarni, A.K.,
ApJ \textbf{398}, 1105 (2009)

\bibitem{RucinskiEtAl2008}  Rucinski, S.M., et al. MNRAS \textbf{391}, 1913 (2008)

\bibitem{vanderKlis2006} van der Klis, M., {\it Compact Stellar X-Ray Sources}, Eds. W.H.G. Lewin
\& M. van der Klis (Cambridge: Cambridge Univ. Press, 2006), p. 39



\bibitem{BelloniEtAl2007} Belloni, T., M\'endez, M., Homan, J., MNRAS \textbf{376}, 1133
(2007)


\bibitem{DonatiEtAl2007} Donati, J.-F., Jardine, M. M., Gregory, S. G., et al.,
MNRAS \textbf{380}, 1297 (2007)


\bibitem{JohnsKrull2007} Johns-Krull, C. M., ApJ \textbf{664}, 975 (2007)


\bibitem{Gregory2011} Gregory, S. G. 2011, American Journal of Physics \textbf{79}, 461



\bibitem{LongEtAl2007} Long, M., Romanova, M.M., \& Lovelace, R.V.E., MNRAS, \textbf{374},
436 (2007)


\bibitem{LongEtAl2008} Long, M., Romanova, M.M., \& Lovelace, R.V.E., MNRAS, \textbf{386},
1274 (2008)


\bibitem{LongEtAl2012} Long, M., Romanova, M.M., Lamb, F.K., New Astronomy
\textbf{17}, 232 (2012)



\bibitem{DonatiEtAl2008} Donati J.-F., Jardine M. M., Gregory S. G. et al., MNRAS
\textbf{386}, 1234 (2008)


\bibitem{RomanovaEtAl2011a} Romanova, M. M., Long, M., Lamb, F. K., Kulkarni, A. K.,
 Donati, J.-F., MNRAS, \textbf{411}, 915 (2011)


\bibitem{LongEtAl2011} Long, M., Romanova, M.M., Kulkarni, A.K., Donati, J.-F.,  MNRAS \textbf{413},
1061 (2011)


\bibitem{AlencarEtAl2012} Alencar S.H.P., Bouvier, J,  Walter, F.M., Dougados, C., Donati, J.F., Kurosawa, R.,
Romanova, M.M. et al., A\&A \textbf{541}, A116 (2012)


\bibitem{Lai1999} Lai, D., ApJ \textbf{524}, 1030 (1999)



\bibitem{LipunovShakura1980} Lipunov, V.M., \& Shakura, N.I., Soviet Astron. Let.
\textbf{6}, 14 (1980)


\bibitem{RomanovaEtAl2013} Romanova, M.M., Ustyugova, G.V., Koldoba, A.V.,
Lovelace, R.V.E., MNRAS \textbf{430}, 699 (2013)


\bibitem{TerquemPapaloizou2000} Terquem, C., \&  Papaloizou, J.C.B., A\&A \textbf{360}, 1031
(2000)

\bibitem{AlencarEtAl2010} Alencar, S.H.P. et al., A\&A \textbf{519}, A88 (2010)

\bibitem{BouvierEtAl1999} Bouvier J. et al., A\&A \textbf{349}, 619 (1999)

\bibitem{LovelaceRomanova2007}  Lovelace, R. V. E. \& Romanova, M. M., ApJ \textbf{670},
L13 (2007)



\bibitem{IllarionovSunyaev1975} Illarionov, A. F., \& Sunyaev, R. A., A\&A, \textbf{39},
185 (1975)

\bibitem{LovelaceEtAl1999} Lovelace, R.V.E., Romanova, M.M., Bisnovatyi-Kogan, G.S.,
ApJ, \textbf{514}, 368 (1999)


\bibitem{RomanovaEtAl2005} Romanova, M.M., Ustyugova, G.V., Koldoba, A.V.,
Lovelace, R.V.E., ApJ \textbf{635}, 165L (2005)


\bibitem{RomanovaEtAl2009} Romanova, M.M., Ustyugova, G.V., Koldoba, A.V.,
Lovelace, R.V.E., MNRAS \textbf{399}, 1802  (2009)


\bibitem{UstyugovaEtAl2006} Ustyugova, G.V., Koldoba, A.V., Romanova, M.M.,
 Lovelace, R.V.E., ApJ \textbf{646}, 304 (2006)


\bibitem{GoodsonEtAl1997} Goodson, A.P., Winglee, R. M., B\"ohm, K.-H., ApJ
\textbf{489}, 199 (1997)


\bibitem{LiiEtAl2013} Lii, P.S., Romanova, M.M., Ustyugova, G.V., Koldoba, A.V., Lovelace, R.V.E.
 MNRAS, in press (arXiv:1304.2703) (2013)


\bibitem{DangeloSpruit2010} D'Angelo C. R., \& Spruit H. C., MNRAS \textbf{406}, 1208
(2010)

\bibitem{LovelaceEtAl2002} Lovelace, R.V.E., Li, H., Koldoba, A.V., Ustyugova, G.V., Romanova,
M.M., ApJ \textbf{572}, 445 (2002)

\bibitem{LovelaceEtAl1991} Lovelace, R.V.E., Berk, H.L., Contopoulos, J., ApJ \textbf{379},
696 (1991)

\bibitem{vanStraatenEtAl2005} van Straaten, S., van der Klis, M., Wijnands, R., ApJ
\textbf{619}, 455 (2005)

\bibitem{PatrunoEtAl2009} Patruno, A., Watts, A.L., Klein-Walt, M., Wijnands, R., van
der Klis, M., ApJ \textbf{707}, 1296

\bibitem{PatrunoDangelo2013} Patruno, A., \& D'Angelo, C., ApJ \textbf{771}, 94 (2013)

\bibitem{Mauche2006} Mauche, C.W., MNRAS \textbf{369}, 1983 (2006)

\bibitem{KurosawaRomanova2012} Kurosawa, R., \& Romanova, M.M., MNRAS \textbf{426},
2901 (2012)


\bibitem{LiiEtAl2012} Lii, P.S., Romanova, M.M., Ustyugova, G.V., Koldoba, A.V., Lovelace, R.V.E.
 MNRAS \textbf{420}, 2020 (2012)


\bibitem{CalvetEtAl1993} Calvet N., Hartmann L., Kenyon S. J., ApJ \textbf{402},
623 (1993)

\bibitem{KoniglEtAl2011} K\"onigl, A., Romanova, M.M., Lovelace, R.V.E., MNRAS
\textbf{416}, 757 (2011)


\bibitem{KurosawaEtAl2011} Kurosawa, R., Romanova, M.M., Harries, T.J.,  MNRAS \textbf{416}, 2623 (2011)


\bibitem{EdwardsEtAl2006} Edwards, S., Fischer, W., Hillenbrand, L., Kwan, J.,
ApJ \textbf{646}, 319 (2006)

\bibitem{BacciottiEtAl1999} Bacciotti, F., Eisloffel, J.,  Ray, T.P., A\&A \textbf{350},
917 (1999)

\bibitem{LovelaceEtAl2010b}  Lovelace, R.V.E., Romanova, M.M., Ustyugova, G.V.,
Koldoba, A.V., MNRAS \textbf{408}, 2083 (2010)



\bibitem{PapittoEtAl2007}
    Papitto, A., di Salvo, T., Burderi, L., Menna, M. T., Lavagetto, G., Riggio, A.,
    MNRAS \textbf{375}, 971 (2007)

\bibitem{AlyKuijpers1990} Aly, J. J., \& Kuijpers, J., A\&A \textbf{227}, 473
(1990)








\end{thebibliography}

\end{document}